\documentclass[12pt,a4paper]{article}
\usepackage{amsmath}
\usepackage{amsfonts}
\usepackage{amssymb}

\author{Andrei Khrennikov and Irina Basieva\\ 
International Center for Mathematical Modeling\\
 in Physics and Cognitive Sciences \\
 Linnaeus University,  V\"axj\"o-Kalmar, Sweden}

\title{Quantum model for psychological measurements: from the projection postulate to 
interference of mental observables represented as positive operator valued measures}

\begin{document}

\maketitle

\begin{abstract} 
Recently foundational issues of applicability of the formalism of quantum mechanics (QM) to cognitive psychology,  decision making, and psychophysics  
attracted a lot of interest. In particular, in \cite{DKBB} the possibility to use of the projection 
postulate and representation of ``mental observables'' by 
Hermitian  operators was discussed in very detail. The main conclusion of the recent 
discussions on the foundations of  ``quantum(-like) cognitive psychology'' is that one has to be 
careful in determination of conditions of applicability of the projection postulate as a mathematical tool
for description of measurements of observables represented by Hermitian operators. 
To represent some statistical experimental data (both physical and mental) in the quantum(-like)
way, one has to use  generalized quantum observables given 
by positive operator-valued measures (POVMs). This paper contains a brief review on POVMs which can be useful 
for newcomers to the field of quantum(-like) studies.  Especially interesting for
cognitive psychology is  a variant of the formula of total probability (FTP) with the interference term derived for 
incompatible observables given by POVMs. We present an interpretation of the interference term from the psychological
viewpoint. As was shown before, the appearance of such a term (perturbing classical FTP) plays the important role in 
cognitive psychology, e.g., recognition of ambiguous   figures and the disjunction effect. The interference term for 
observables given by POVMs has much more complicated structure than the corresponding term for observables given by 
Hermitian operators. We elaborate cognitive interpretations of different components of the POVMs-interference term
and apply our analysis to a quantum(-like) model of decision making.  
 \end{abstract}

\section{Introduction}

Recently the mathematical formalism of QM started to be widely used in various domains outside 
of physics, see, e.g., \cite{AS1}--\cite{W1}.
We are interested in applications to cognitive psychology,
especially decision theory \cite{AS3}--\cite{AS7}, \cite{Jerome1}--\cite{Jerome3},  \cite{Cheon1}--\cite{Cheon2}, 
\cite{K5}--\cite{K7}, \cite{POT}--\cite{POT2}, and psychophysics \cite{Dzhafarov}.
There is a plenty of statistical data from experiments on recognition of ambiguous figures \cite{Conte1}--\cite{Conte3}, 
the disjunction effect (as well as other probability judgment fallacies) \cite{T1}, \cite{T2}, 
\cite{Jerome1}--\cite{Jerome5},  and 
psychophysics \cite{Dzhafarov},  violating the classical formula of 
total probability (FTP). It was well known, see, e.g., Feynman
  et al. \cite{Feynman}, that statistical data obtained in quantum experiments 
of the interference type, e.g., the two slit experiment, violate FTP, see \cite{KINT} for detailed analysis. 
The additional interference term in the quantum version of FTP can be naturally expressed through the wave function and 
eigenvectors of Hermitian operators representing 
observables with the aid of  the projection postulate. The latter describes 
measurements of quantum observables (represented by Hermitian operators).

Since FTP is violated for data from both QM and cognitive psychology and the former developed the advanced mathematical 
apparatus for study of the interference effect, it is natural to apply this apparatus to describe statistical data from the latter,
see above references and especially Khrennikov \cite{K2}, \cite{K3}, \cite{K7}, Busemeyer et al. \cite{Jerome1}, Dzhafarov et al. \cite{Dzhafarov}.     
However, there is one pitfall; 
one problem has to be taken into account and carefully analyzed. The experimentally obtained  matrices of  
transition probabilities\footnote{The elements 
of such a matrix are  probabilities of transition from one state to another as the result of measurement. Matrices of transition probabilities
of QM can be considered as analogs  of matrices of transition probabilities in theory of classical Markov systems. 
(However, one has to be careful when exploring such an analogy.)}  are not {\it doubly stochastic} \cite{K5}, \cite{K7}, 
 \cite{H2} -- as they have to be for transitions from one orthonormal basis 
to another.  Here we discuss the bases of 
the eigenvectors of the Hermitian operators representing quantum observables.  Therefore ``cognitive observables'' cannot be 
represented
by Hermitian operators (at least with nondegenerate spectra).  In quantum physics the same problem was resolved by invention of 
generalized quantum observables, {\it positive operator-valued measures} (POVMs). In general  the matrices of transition probabilities
for measurements represented by POVMs are not doubly stochastic. Therefore it is natural to apply POVMs to describe the 
aforementioned  statistical data. 
In this paper we proceed in this  direction. 

We remark that  in cognitive psychology, decision making, and psychophysics there are other important reasons to proceed with generalized quantum observables given by POVMs 
(and not with the Dirac-von Neumann observables
given by Hermitian operators), see \cite{DKBB} for detailed analysis.    
However, in this paper we concentrate mainly on violation of double stochasticity for statistical data with nonclassical features, 
e.g., in the form of violation of FTP collected in experiments in cognitive 
psychology. 

We start with analysis of possible sources of violation of double stochasticity and point to violation of the  projection 
postulate as the key point, see section \ref{PPP} on a general discussion on applicability of this postulate in  physics,
cognitive science, psychology, and psychophysics. 

As an example illustrating  possible applications of POVMs, 
we present a quantum-like  model of decision making based on representation of 
the {\it  decision and intention observables} by POVMs.
Here (as well as in models based on the standard representation of observables by Hermitian 
operators) the violation of classical FTP is of the fundamental value.  

Although the inter-relation between violation of the classical FTP and quantum interference of probabilities  
is well studied \cite{Feynman}, \cite{KINT}, 
typically the situation was restricted to probabilities corresponding to the conventional quantum observables and the pure 
states (Hermitian operators
and complex state vectors). The FTP with interference 
term for the POVM-observables and the mixed states (density operators) was derived  in   \cite{LOUB}. 
In this case the interference term cannot be parametrized just by the ``interference angle'' (phase) $\theta.$ 
It contains a few parameters 
having interesting cognitive interpretations, see section \ref{DM}. Here we present general theoretical analysis 
of the structure of the POVM-interference. However, the real cognitive meaning will become clear only in future through 
study of various concrete examples.   

We also point out that the POVMs approach dismiss the exceptional role of {\it pure quantum states} 
(which are mathematically represented by normalized vectors in complex Hilbert space). This approach 
is based on usage of general quantum states given by {\it density operators}\footnote{These are 
Hermitian, positively defined and trace one operators.} which represent in general classical
statistical mixtures of superpositions.

The quantum-like model of decision making in the two party games
based on usage of pure states (to represent mental states), Hermitian operators (to represent 
decision and intention observables), and the projection postulate (to represent measurements) 
\cite{Jerome1}--\cite{Jerome3},  \cite{Cheon1}--\cite{Cheon2}, 
\cite{K5}--\cite{K7}, \cite{POT}--\cite{POT2} can be generalized to the model with density operators, 
POVMs, and the corresponding formulas for measurement output probabilities and states. 

One party, say Bob, has to make decision 
on some problem $B.$ The payoff depends on the actions of another party, say Alice.  Therefore 
Bob has to take into account intentions $A$ of Alice. The mental state of Bob is described
by the density operator $\rho,$ his decision observable is given by the POVM $M^{b};$ Bob also 
has to analyze possible actions of Alice, output of this analysis is treated as 
(self-)measurement and it  is represented by the POVM 
$M^{a}.$ 

We emphasize that both the decision making observable $M^{b}$ and the intention observable 
$M^{a}$ represent {\it self-measurements.} This is the crucial interpretational difference from 
quantum measurement theory. In the latter, an observer is sharply separated from 
a system. The simplest resolution of the ``self-measurement problem'' is through consideration 
of the brain as a complex system unifying numeorus subsystems having some degree of independence.
We might assume that some subsystems process the mental states and other subsystems perform measurements.
However, we understood well that the problem is of extreme complexity. In fact, this is the problem 
of transition from unconsciousness to consciousness. We shall not try to go deeper into this problem and 
we shall proceed at the  formal operational level, cf. \cite{BUS}--\cite{DA}. 
  
In section \ref{GPP} we start with a brief introduction to theory of generalized quantum observables, POVMs. 
Then we present the FTP with interference
term corresponding to the most general quantum interference: a quantum state given by a density operator  
and incompatible generalized observables 
given by a pair of  POVMs, see \cite{LOUB}. Finally, we proceed to a quantum-like model of decision making based on POVMs.
(We call our model quantum-like to distinguish it from really quantum physical models which were use by some authors to model 
information processing by the brain, e.g., \cite{P1}--\cite{HA2}. Our model has no relation to quantum physical carriers of 
information.) 

Finally, we remark that quantum measurement theory is far from 
being satisfactory (and this is well recognized by physicists). It might happen that 
the present experiments with extension of its domain of applications 
to cognitive psychology, decision making, and psychophysics would clarify  some of its problems.  

\section{Projection postulate in physics and psychology}
\label{PPP}

\subsection{Projection postulate in quantum physics}

For readers convenience, we recall this postulate of QM; for simplicity we consider the finite dimensional case.  

\medskip

Suppose that the state  of a quantum system is represented by the (norm-one) vector $\psi$  of  
complex Hilbert space and suppose that 
a quantum observable is represented by the Hermitian operator $A.$ 
If measurement of this observable is resulted in the concrete eigenvalue $a$ of $A$ (and by the spectral postulate
of QM only eigenvalues can be observed), then the state-vector $\psi$ is projected onto the eigensubspace
of $H$ corresponding to $a.$  (Since the projection can have the norm less than one, it has to be renormalized by dividing 
by its norm.\footnote{Thus the operation of measurement is nonlinear -- opposite to the Schr\"odinger dynamics of an
 isolated quantum system.})

In the case of nondegenerate spectrum all eigensubspaces are one-dimensional and the aforementioned projection (after renormalization) is just 
the eigenvector $e_a$ of $A$ corresponding to $a.$ One typically says that the state $\psi$ collapsed to the eigenstate $e_a.$ In the case
of continuous variables (with representation in the infinite-dimensional Hilbert space $L_2$ of square integrable functions) states are given
by complex valued functions $\psi(x).$ Here one says about collapse of the wave function as the result of measurement.  

\medskip


This postulate plays a special role in axiomatics of QM. Although majority of quantum 
community uses it routinely and operates with the notion of collapse of the wave function, 
a part of this community handle this postulate with suspicion.
This postulate remains controversial, see Auyang \cite{Auyang}, p. 23, for the detailed discussion. 
In particular, E. Beltrametti and G. Cassinelli (who are  among the world leading experts in quantum logic and foundations) 
remarked \cite{BELT} that ``it does not have the status of postulates of quantum theory,  necessary for its internal 
coherence.'' 

L. E. Ballentine (who is also one of the world leading experts in quantum foundations, 
one of creators of the ``statistical interpretation of QM'') 
pointed out \cite{BL} that this postulates leads to wrong conclusions. Even if a 
quantum system through interaction with a measurement device triggers it to produce one fixed 
eigenvalue, in general the state of this system does not collapse. As an example of inapplicability 
of the projection postulate, he considered the track left by a quantum (charged) particle
 in a cloud chamber. Typically the state of the incoming particle is given by a momentum amplitude.
 Particles ionizes the fist atom in chamber and this process is considered as position measurement.
 By the projection postulate particles state should collapse to the corresponding eigenstate of 
the position operator, but the latter is a spherical wave which spreads out uniformly in all 
directions. Hence, it would be impossible for for this particle to ionize subsequent atoms to form 
a track which indicates the direction of original momentum, which is allegedly destroyed by the 
first ionization, see \cite{BL}.

 One of coauthors of this paper \cite{K_EPR}, \cite{K_EPR1} analyzed the role 
of the projection postulate in reasoning on incompleteness/non-locality of QM which was presented
in the famous argument of Einstein, Podolsky, and Rosen, known as the EPR paradox. In fact, this paradox 
can be reduced to the use of the projection postulate \cite{K_EPR}, \cite{K_EPR1}. 

Finally, we remark that the projection postulate in the presently common 
form of the projection of the state vector (``collapse of the wave function'') was proposed by J. von Neumann only for
observables with {\it nondegenerate spectra} \cite{VN}. It was extended to all quantum observables by L\"uders, see, e.g.,  \cite{BUS} for details. 
However, from the very beginning of QM
majority of physicists applied this postulate without paying attention to the form of spectra of observables, 
e.g., in the aforementioned EPR argument. At the same time von Neumann pointed out that one must differ sharply
the cases of nondegenerate and degenerate spectra. In the latter case he elaborated a more complicated mathematical scheme than
simply vector reduction. (Thus von Neumann did not accept ``collapse on eigensubspace'' of the dimension larger than one.) 

In physics understanding of the ``state projection problem'' in 1970th generated a constructive reply in the form of generalization of the class
of quantum observables and, hence, the mathematical scheme of quantum measurement. The first step in this direction was consideration of
quantum observables represented not by Hermitian operators, but by POVMs. The main idea beyond this generalization is very simple. Any Hermitian
operator can be characterized by its spectral family. In the finite dimensional case this is just the family of  projectors to its eigensubspaces.
These projectors are mutually orthogonal.
Any projector is by itself Hermitian (self-adjoint), i.e., $P^* = P,$ idempotent, i.e., $P^2 =P,$  positively defined, i.e., 
$\langle P\psi, \psi\rangle \geq 0,$ for any vector $\psi,$ and the
projectors from the spectral family sum up to the unit operator, in the case of discrete spectral family $(P_n)$ we have
$$
\sum_n P_n =I,
$$
where $I$ is the unit operator in the Hilbert state space. In the definition of POVMs operators need not be idempotent, i.e., 
these are Hermitian positively defined operators.    

\subsection{Projection postulate in cognitive psychology and decision making}

In applications of quantum theory to cognitive psychology, decision making, and psychophysics  the role of the projection postulate 
(in the L\"uders form) was analyzed in very detail in the paper \cite{DKBB} (see also \cite{K7}, \cite{H2}). The authors of \cite{DKBB} elaborated 
examples of mental measurements similar to the  Ballentine example with a charged particle interacting with ions in a cloud chamber. In combination with aforementioned problem of violation of 
double stochasticity of the matrix of transition probabilities the examples  from \cite{DKBB} motivate the use of POVMs, instead of Hermitian
operators. (However, the authors of the paper \cite{DKBB} also found
that there exist problems of decision making which seems to be impossible to describe operationally 
even with the aid of POVMs, cf. \cite{B7}.)

\section{Neumark dilation theorem and theory of open quantum systems in psychology}

Another possibility to solve the problem of {\it non-double stochasticity of matrices of transition probabilities for 
data on recognition of ambiguous figures and the disjucntion effect} is to consider Hermitian operators    
with degenerate spectra acting in Hilbert space having the dimension which is higher than the ``natural dimension'' of the decision
problem. In the simplest case, the decision and intention  observables are dichotomous and the ``natural Hilbert space 
of the game'' is two  dimensional, i.e., observables are represented by $2\times 2$ matrices.      
We state again that it is impossible to represent some psychological statistical data by Hermitian $2\times 2$ 
matrices. If, instead of using POVMs, one  considers Hermitian operators in Hilbert spaces of higher 
dimensions, the problem of the interpretation of these extra dimesnions arises.  At the present   time neurophysiology cannot 
present the complete set of variables involved in the decision making. Therefore we do not know what and how many mental 
variables have to be added. The ``neurophysiological dimension'' of the Hilbert space of mental states   may be millions or even billions. 
Therefore a possibility to solve the problem of non-double stochasticity by using $2\times 2$ POVMs, instead of 
$\rm{billion} \times  \rm{billion}$ Hermitian matrices, is more attractive. 

We recall that by {\it Neumark's theorem} \cite{NEU}
any generalized quantum observable given by a POVM $M$ can be represented as a conventional quantum observable  given by 
a Hermitian operator $A$ acting in the Hilbert space of higher dimension. This extended Hilbert space can be represented as the tensor 
product of the original Hilbert space $H$ and another Hilbert state space $K.$ Here $A$ acts in $H\otimes K.$ It can be considered 
as a measurement on a compound system, the original system $S$ and  another system $S^\prime.$ An adequate mental interpretation 
of such lifting of (self-)measurement on $S$ to (self-)measurement on a compound mental system $(S, S^\prime)$ has not yet been 
elaborated. Representations of measurement by POVM $M$ acting in $H$ and Hermitian operator $A$ acting in $H\otimes K$ are coupled
by an isometric operator, say $B.$ Its cognitive meaning is neither clear. 

We point out  that in physics appearance of POVMs and  the Neumark theorem are typically treated in the 
framework of theory of open quantum systems. Additional degrees of freedom represented by the Hilbert state space $K$ 
in the above consideration are interpreted as representing an environment, in our case, so to say, the {\it mental environment.}
In physics the situation is essentially simpler (from the interpretational viewpoint): one can identify a quantum system as a physical 
entity sharply separated from the environment; for example, an electron under influence of the electromagnetic fluctuations. In cognitive science,
as was already emphasized in introduction, it is not easy to extract physically a cognitive system, as a decision maker, from its 
mental environment. The brain is at the same time a system and observer, self-observer, and more generally it is at the same a system and its 
mental environment, e.g., in the form of memory.

Therefore it is natural to distance from the physical structure of the aforementioned entities and proceed at the purely information level.
Both a decision maker and its environment are considered as information systems exchanging information. The decision is coming as the result 
of such an information exchange which resolves the information uncertainty in a decision maker.    Since the mental (as well as physical) 
environment has huge complexity, one does not try to describe explicitly the compound system, the decision maker plus the environment.
The operational description based on POVMs is explored. In principle, one can proceed at a deeper level of theory of open quantum systems
and describe the dynamics of the state of a decision maker interacting with the mental environment, see Asano {\it et al.} \cite{AS4}--\cite{OH4} .

\section{Mathematical structure of projection postulate}
\label{MMM}

\subsection{States and observables}

\label{DENSITY}

Starting from Heisenberg, observables are represented by Hermitian matrices or
in the abstract framework by Hermitian operators%
\index{Hermitian operator}%
. These operators act in complex Hilbert space $H$, i.e., a complex linear
space endowed with a scalar product denoted as $\langle\psi_{1}\vert\psi
_{2}\rangle.$\footnote{The latter is a function from the Cartesian product
$H\times H$ to the field of complex numbers $\mathbb{C},$ $\psi_{1}, \psi_{2}
\to\langle\psi_{1}\vert\psi_{2}\rangle,$ having the following properties:
\par
1. Positive definiteness: $\langle\psi\vert\psi\rangle\geq0$ with $\langle
\psi\vert\psi\rangle=0$ if and only if $\psi=0.$
\par
2. Conjugate symmetry: $\langle\psi_{1}\vert\psi_{2}\rangle=\overline
{\langle\psi_{2}\vert\psi_{1}\rangle}$
\par
3. Linearity with respect to the first argument: $\langle k_{1} \psi_{1} +
k_{2} \psi_{2}\vert\phi\rangle= k_{1} \langle\psi_{1}\vert\phi\rangle+ k_{2}
\langle\psi_{2}\vert\phi\rangle,$ where $k_{1}, k_{2}$ are complex numbers.
\par
Here bar denotes complex conjugation%
\index{complex conjugation}%
; for $z=x+iy,$ $\bar{z}=x-iy.$ From the second and third properties it is
easy to obtain that, for the second argument, $\langle\phi\vert k_{1} \psi_{1}
+ k_{2} \psi_{2} \rangle=\bar{k}_{1} \langle\phi\vert\psi_{1} \rangle+ \bar
{k}_{2} \langle\phi\vert\psi_{2} \rangle.$ By fixing in $H$ an orthonormal
basis $(e_{j}),$ i.e., $\langle e_{i}\vert e_{j}\rangle= \delta_{ij},$ we
represent vectors by their coordinates $\psi_{1}=(z_{1},...,z_{n},...),
\psi_{2}= (w_{1},...,w_{n},...).$ In the coordinate representation the scalar
product has the form $\langle\psi_{1}\vert\psi_{2}\rangle= \sum_{j} z_{j}
\bar{w}_{j}.$ By using this representation the reader can easily verify the
aforementioned properties of the scalar product.
\par
We remark that usage of \textit{complex numbers}%
\index{complex numbers}
plays the crucial role. One cannot proceed with real Hilbert spaces. There are
experimental statistical data which cannot be embedded in the real model.}

The norm (the abstract analog of the Euclidean length) of a vector is defined
as $\Vert\psi\Vert= \sqrt{\langle\psi\vert\psi\rangle}.$ In the fixed system
of coordinates $\Vert\psi\Vert= \sqrt{\sum_{j} \vert z_{j}\vert^{2}}.$
Normalized vectors of $H,$ i.e., $\psi$ such that $\Vert\psi\Vert=1,$
represent a special (and the most important) class of states of quantum
systems, \textit{pure states.}%
\index{pure state}
Each pure state $\psi$ can be represented as an operator acting in $H,$
namely, the orthogonal projector on this vector. Thus, by fixing the
orthonormal basis in $H,$ we represent any pure state by a matrix $\rho
=(\rho_{ij}):$

\medskip

a) it is positively defined\footnote{Positive definiteness means that, for any
vector $\phi,\; \langle\rho\phi\vert\phi\rangle\geq0.$},

b) Hermitian\footnote{Hermitianity means that $\rho_{ij}= \bar{\rho}_{ij},$ in
particular, the diagonal elements are real.},

c) its trace equals 1.\footnote{Thus $\mathrm{{Tr} \rho= \sum_{j} \rho
_{jj}=1.}$}

\medskip

In the two dimensional case, one quantum bit (qubit%
\index{qubit}%
) case, we have
\begin{equation}
\rho=\left(
\begin{array}
[c]{cc}%
\rho_{11} & \rho_{12}\\
\rho_{21} & \rho_{22}%
\end{array}
\right)  .
\end{equation}
In the two dimensional case positive definiteness means that $\rho_{ii} >0,
i=1,2$ and the determinate is also positive; finally, $\mathrm{{tr} \rho=
\rho_{11} + \rho_{22}=1.}$

The condition guaranteeing that $\rho$ is a projector can be written as the
equality $\rho^{2}= \rho.$ The next step in development of the quantum
formalism (due to Landau and von Neumann) was proceeding without the latter
constraint, i.e., considering all possible matrices satisfying conditions
a)-c). They are called \textit{density matrices} and they represent the most
general states of quantum systems. In the abstract framework one considers
operators satisfying conditions a)-c), density operators%
\index{density operators}%
. Each density operator can be written as a weighted sum of projection
operators corresponding to pure states. If such a sum contains more than one
element, then the state represented by this density operator is called a
\textit{mixed state}%
\index{mixed state}%
: the mixture of pure states with some weights. Although this terminology is
widely used, it is ambiguous. Representation of a density operator as a
weighted sum of projectors corresponding to pure states is not unique. Thus by
using the terminology mixed state one has to take into account this non-uniqueness.

For simplicity let us restrict consideration to the case of finite dimensional
Hilbert space. Take an arbitrary quantum observable, represented by a
Hermitian operator $A.$ For any Hermitian operator, there exists an orthogonal
basis of $H$ consisting of its eigenvectors. The values of the observable
which is operationally represented by $A$ are encoded in the eigenvalues of
this operator, $a_{1},...,a_{n}.$ (We shall use the same symbol for an
observable and its operator representation.)

Consider now the case of nondegenerate spectrum. Here all eigenvalues are nondegenerate, i.e., 
all eigensubspaces are 
one dimensional.
For an ensemble of quantum systems prepared in the same state which is
represented by the density operator $\rho,$ the probability $p(a_{j})$ to obtain the
fixed (eigen)value $a_{j}$ is encoded in the corresponding matrix element of the
operator $\rho$ (in the basis of eigenvectors):
$p(a_{j})= \rho_{jj}.$
This rule can be also written as
\begin{equation}
\label{OOO77}
p(a_{j})=  \rm{Tr} \rho P_{j},
\end{equation}
where $P_j$ is the projector to the eigenvector $e_j$ corresponding the 
eigenvalue $a_j.$ This is one of the basic postulate of quantum mechanics. It connects
experimental probabilities with operator representation of observables. (In
the simplest form, for the pure state, this postulate was proposed by M. Born.

If the spectrum of $A$ is degenerate, i.e., eigensubspaces are in general multi-dimensional,
then the same rule (\ref{OOO77}) for calculation of probabilities is applicable with $P_j$ as
the projector to the eigen-subspace $L_j$ corresponding the 
eigenvalue $a_j.$ 

\subsection{Superposition}
\label{SUPRSUPER}

Let the state space of some system (physical or cognitive) be represented by
finite-dimensional Hilbert space $H.$ Consider the pure state $\psi$ and the
observable $A,$ denote its eigenvalues by $a_{1},..,a_{m}$ and the
corresponding eigenvectors by $e_{1},..., e_{m}.$ This is an orthonormal basis
in $H.$ (We again proceed under the assumption that spectrum is nondegenerate.) 
We expand the vector $\psi$ with respect to this basis:
\begin{equation}
\label{SUPER}\psi= c_{1} e_{1} + ...+ c_{m} e_{m},
\end{equation}
where $(c_{j})$ are complex numbers such that the sum of their squared
absolute values equals to one (this is the coordinate expression of the
normalization by one of a pure state-vector):
$\vert c_{1}\vert^{2} + ...+ \vert c_{m}\vert^{2} =1.
$
By using the terminology of linear algebra we say that the pure state $\psi$
is \textit{superposition}%
of pure states $e_{j}.$
The density matrix corresponding to $\psi$ has the elements
$\rho_{ij}= c_{i} \bar{c}_{j}.
$
Hence, for the pure state $\psi,$ the basic probabilistic postulate of quantum
mechanics, (\ref{OOO77}), has the form:
$
p(a_{j})= \rho_{jj}= c_{j} \bar{c}_{j}= \vert c_{j}\vert^{2}.
$
This postulate can be written without using the coordinates of the state
vector $\psi$ with respect to the basis of eigenvectors of a quantum
observable. We remark that, since the basis of eigenvectors of a Hermitian
operator can always be selected as orthonormal, the coordinates $c_{j}$ can be
expressed in the form: $c_{j}= \langle\psi\vert e_{j}\rangle.$ Hence, the
Born's rule takes the form:
\begin{equation}
\label{OOO177}p(a_{j})= \vert\langle\psi\vert e_{j}\rangle\vert^{2}.
\end{equation}

\subsubsection{Projection postulate; resolution of uncertainty}

Consider the case of nondegenerate spectrum.

\label{PROJECTION}

The next natural question is about the post-measurement state. What will
happen with the state $\psi$ after measurement? By the \textit{projection postulate}%
the superposition (\ref{SUPER}) is reduced to just one term, the state $e_{j}$
corresponding to the eigenvalue $a_{j}$ which was obtained in the measurement.
This procedure can be interpreted in the following way:

\medskip

\textit{This superposition encodes uncertainty in results of measurements for
the observable $A.$ Roughly speaking before measurement, a quantum system
``does not know how it will answer to the question $A.^{\prime\prime}$ The
mathematical expression (\ref{SUPER}) encodes potentialities
for different answers. Thus a quantum system in the superposition state $\psi$
does not have any value of $A$ as its objective property
. After measurement superposition is
reduced to just one term in the expansion (\ref{SUPER}) corresponding the
value of $A$ obtained in the process of measurement.}

\medskip

We remark that the state reduction is often called state's \textit{collapse.}%
\index{collapse}
Some experts in quantum foundations treat superposition physically and not
simply operationally; for them, the collapse is also the physical event. We
proceed with the operational interpretation
of the quantum formalism. By this interpretation superposition (\ref{SUPER})
express uncertainty in expected results of the $A$-measurement. 
 When the result $a_{j}$ is detected, this uncertainty is resolved.
Hence, ``collapse'' takes place not in physical space, but in information space.

Encoding of uncertainty in cognitive systems by superpositions is one of the
cornerstones of quantum(-like) approach to cognition. A cognitive system as well as a quantum
physical system can be in a state of uncertainty on possible reactions to
measurements. Such states are \textit{mathematically encoded} as linear
superpositions. Measurement resolves such superpositions.

This picture of resolution of information uncertainty in the state of a cognitive system
is very clear and simple and therefore it is very attractive. It can be applied to some 
measurements performed in cognitive science and psychology. It is also useful for presentation 
of the fundamentals of quantum measurement theory for researchers working in cognitive science
and psychology \cite{K7}, \cite{Jerome5}, \cite{K2}. In fact, the same happens in quantum physics: the majority of textbooks
on quantum mechanics are based on this picture of measurement, only more advanced books, especially with applications
to quantum information theory, present a more complex picture of resolution of quantum uncertainty. 

In real experiments the situation is more complicated. It is not so often possible to resolve uncertainty completely 
as the result of measurement. Typically uncertainty is reduced only partially and the output of measurement cannot be presented
by the projection postulate. As was emphasized in introduction, more general observables, POVMs, and corresponding state updating scheme,
see section \ref{PP}, have to be explored. The heuristic picture of POVM-measurement is that this is a kind of fuzzy measurement, see especially
\cite{BUS}, \cite{DMU}. For a fixed input pure state, the output states corresponding to different eigenvalues are not mutually orthogonal,
as it is in the case of observables represented by Hermitian operators. For dichotomous decision observable $A= 
0,1,$ its values cannot be be interpreted as sharp ``no'', ``yes''. The corresponding output subspaces are in general overlaping;
Roughly speaking such ``no''-decision is partially ``yes'' and vice versa. A proper cognitive interpretation has to be elaborated.                                                                                                                                                                                                                                                                                                                                                                                                                                                                                                                                                                                                                                                                                                                                                                                                                                                                                                                                                                                                                                                                                                                                                                                                                                                                                                                                                                                                                                                                                                                                                                                                                                                                                                                                                                                                                                                                                                                                                                                                                                                                                                                                                                                                                                                                                                                                                                                                                                                                                                                                                                                                                                                                                                                                                                                                                                                                                                                                                                                                                                                                                                                                                                                                                                                                                                                                                                                                                                                                                                                                                                                                                                                                                                                                                                                                                                                                                                                                                                                                                                                                                                                                                                                                                                                                                                                                                                                                                                                                                                                                                                                                                                                                                                                                                                                                                                                                                                                                                                                                                                                                                                                                                                                                                                                                                                                                                                                                                                                                                                                                                                                                                                                                                                                                                                                                                                                                                                                                                                                                                                                                                                                                                                                                                                                                                                                                                                                                                                                                                                                                                                                                                                                                                                                                                                                                                                                                                                                                                                                                                                                                                                                                                                                                                                                                                                                                                                                                                                                                                                                                                                                                                                                                                                                                                                                                                                                                                                                                                                                                                                                                                                                              

\section{Generalized quantum observables}
\label{GPP}

\subsection{POVMs}
\label{PP}

{\bf Definition.} {\it A positive operator valued measure (POVM) is a family of positive operators $\{M_j\}$ such that 
$\sum_{j=1}^m M_j=I,$ where $I$ is the unit operator.}

\medskip

(We considered the simplest case of a discrete operator valued measure on the set of indexes $J=\{1,2,..., m\},$ see, e.g., \cite{BUS} for 
POVMs on continuous sets.)   It is convenient to use the following representation of POVMs:
\begin{equation}
\label{EQR0}
M_j= V_j^\star V_j,
\end{equation}
 where $V_j: H \to H$ are linear operators.
 
A POVM can be considered as a random observable. Take any set of labels 
$\alpha_1,..., \alpha_m,$ e.g., for $m=2, \alpha_1= \rm{yes}, \alpha_2= \rm{no}.$ Then the corresponding observable 
takes  these values (for systems in the state $\rho)$  with the probabilities 
\begin{equation}
\label{EQR}
p(\alpha_j) \equiv p_\rho(\alpha_j)= \rm{Tr} \rho M_j = \rm{Tr} V_j \rho V_j^\star.
\end{equation}

We are also interested in the {\it post-measurement states.} Let the state $\rho$ was given, a generalized observable 
was measured and the value $\alpha_j$ was obtained. Then the output state  after this measurement 
has the form 
\begin{equation}
\label{EQR1}
\rho_j = \frac{V_j \rho V_j^\star}{ \rm{Tr}V_j \rho V_j^\star}.
\end{equation}

\subsection{Interference of probabilities  for generalized observables}

Consider two generalized observables $a$ and $b$ corresponding to POVMs $M^{a}=\{V_j^\star V_j\}$ and 
$M^{b}=\{W_j^\star W_j\},$ where $V_j\equiv V(\alpha_j)$ and $W_j=W(\beta_j)$  correspond to the values $\alpha_j$ and $\beta_j.$  

If there is given the state $\rho$ the probabilities of observations of values $\alpha_j$ and $\beta_j$ have the form
\begin{equation}
\label{EQR2}
p^a(\alpha)= \rm{Tr} \rho M^{a}(\alpha)= \rm{Tr} V(\alpha)  \rho V^\star(\alpha), \; p(\beta)= \rm{Tr} \rho M^{b}(\beta)
= \rm{Tr} W(\beta) \rho W^\star(\beta).
\end{equation}

Now we consider two consequtive measurements: first the $a$-measurement and then the $b$-measurement. 
If in the first measurement the value $a=\alpha$ was obtained, then the initial state $\rho$ was tranformed
into the state
\begin{equation}
\label{EQR3}
\rho^a_\alpha =  \frac{V(\alpha) \rho V^\star(\alpha)}{ \rm{Tr}V(\alpha) \rho V^\star(\alpha)}.
\end{equation}
For the consequtive $b$-measurement, the probability to obtain the value $b= \beta$ is given by 
\begin{equation}
\label{EQR4}
p(\beta\vert \alpha)= \rm{Tr} \rho^a(\alpha) M^{b}(\beta) =
\frac{\rm{Tr} W(\beta) V(\alpha) \rho V^\star(\alpha)W^\star(\beta)}{ \rm{Tr}V(\alpha) \rho V^\star(\alpha)}.
 \end{equation}
 This is the conditional probability to obtain the result $b=\beta$ under the condition of the result $a=\alpha.$ 
 
 We set 
\begin{equation}
\label{EQR5}
 p(\alpha, \beta) = p^a(\alpha) p(\beta\vert \alpha).
\end{equation}
This is the probability to obtain the result $(\alpha, \beta)$ in the consequtive measurement 
of $a$ and then $b.$ To find $p(\alpha, \beta)$ with the aid of the probability   $p(\alpha)$
and the conditional probability $p(\beta\vert \alpha),$ we apply the classical Bayes formula. 
Thus ``quantumness'' is in the probabilities $ p(\alpha)$ and $p(\beta\vert \alpha);$
the probability $p(\alpha, \beta)$ is the result of their classical combination.

We remark that the probability $p(\alpha, \beta)$ can be expressed directly in terms of POVMs:
\begin{equation}
\label{EQR5a}
p(\alpha, \beta)= \rm{Tr}  \rho V^\star(\alpha)  W^\star(\beta) W(\beta) V(\alpha).
\end{equation}

In general, the family of probabilities $p(\alpha, \beta)$ cannot be considered 
as the joint probability distribution of two classical random variables, the pair $(a, b).$ 
In the same way we introduce  the pobability 
\begin{equation}
\label{EQR6}
 p(\beta, \alpha) = p^b(\beta)p(\alpha \vert  \beta).
\end{equation}
This is the probability to obtain the result $b=\beta$  in the first measurement and then 
the result $a=\alpha.$ One can easily find examples of POVMs (and even Hermitian 
operators), see, e.g., \cite{K7},  such that  
\begin{equation}
\label{EQR7}
p(\alpha, \beta) \not= p(\beta, \alpha).
 \end{equation}

We call $p(\alpha, \beta)$ and  $p(\beta, \alpha)$ {\it ordered joint probabilities.}

 \section{Formula of total probability with the interference term}
 
 We recall that, for two classical random variables $a$ and $b$ which can be represented 
 in the Kolmogorov measure-theoretic approach, the formula of total probability (FTP) has the form
\begin{equation}
\label{EQR8}
 p^b(\beta)= \sum_\alpha p^a(\alpha) p(\beta \vert \alpha).
 \end{equation}
 Further we restrict our consideration to the case of dichotomous variables, $\alpha=\alpha_1, \alpha_2$ and 
 $\beta= \beta_1, \beta_2.$

 This formula which is one of the most fundamental and successful in applications (the  law of total probability)  
 is violated  in quantum physics.  
For example, in the famous two-slit experiment, the probability that a photon is detected at position $x$
  on the photo-sensitive plate (used for measurement) is represented as  
\begin{eqnarray*}
p(x) &=&\left\vert \frac{1}{\sqrt{2}}\psi _{1}(x)+\frac{1}{\sqrt{2}}\psi
_{2}(x)\right\vert ^{2} \\
&=&\frac{1}{2}\left\vert \psi _{1}(x)\right\vert ^{2}+\frac{1}{2}\left\vert
\psi _{2}(x)\right\vert ^{2}+\left\vert \psi _{1}(x)\right\vert \left\vert
\psi _{2}(x)\right\vert \cos \theta, 
\end{eqnarray*}%
where $\psi_1$ and $\psi_2$ are two wave functions, whose absolute values $\left\vert \psi _{k}(x)\right\vert ^{2}$ give
the distribution of photons which pass through the slit-$k$ ($k=0,1$). 
The term of $\left\vert \psi _{1}(x)\right\vert \left\vert \psi_{2}(x)\right\vert \cos \theta $ 
implies the interference effect of two wave functions. 
Let us denote $\left\vert \psi _{k}(x)\right\vert ^{2}$ by $p(x|k)$, and then
the above equation is represented as 
\begin{equation}
\label{AS}
p(x)=p(1) p(x|1)+p(2) p(x|2)+2\sqrt{p(1)p(x|1)p(2)p(x|2)}\cos \theta,
\end{equation}%
where $p(1)=p(2)=1/2.$ In the above form, it seems
that a classical probability law (\ref{EQR8})
is violated, and the term of interference $2\sqrt{p(1) p(x|1)(2) p(2) p(x|2)p}\cos \theta$ specifies the violation.\par

We are now interested in the version of FTP with the interference term for in general nonpure states given by density 
operators and generalized quantum observables given by two (dichotomous)  PVOMs:
\begin{equation}
\label{EQR9}
p^b(\beta) = p^a(\alpha_1) p(\beta\vert \alpha_1) + 
p^a(\alpha_2) p(\beta\vert \alpha_2) + 2 \lambda_\beta \sqrt{p^a(\alpha_1) p(\beta\vert \alpha_1)p^a(\alpha_2) p(\beta\vert \alpha_2)},
\end{equation}
or by using ordered joint probabilities
\begin{equation}
\label{EQR9_a}
p^b(\beta) = p(\alpha_1, \beta) + 
p(\alpha_2, \beta) + 2 \lambda_\beta \sqrt{p(\alpha_1, \beta)p(\alpha_2, \beta)}.
\end{equation}
Here the coefficient of interference $\lambda_\beta$ has the form:
\begin{equation}
\label{EQR10}
 \lambda_\beta= \frac{\sum_{i=1,2} \rm{Tr} \rho \{W^\star(\beta)V^\star(\alpha_i) V(\alpha_i)W(\beta) -
V^\star(\alpha_i) W^\star(\beta)W(\beta)V(\alpha_i) \}}{2 \sqrt{p^a(\alpha_1) p(\beta\vert \alpha_1)p^a(\alpha_2) p(\beta\vert \alpha_2)}}
\end{equation}
Introduce the parameters
\begin{equation}
\label{EQR11}
\gamma_{ \alpha \beta} = \frac{\rm{Tr}  \rho W^\star(\beta)V^\star(\alpha) V(\alpha)W(\beta)}{\rm{Tr} \rho V^\star(\alpha) W^\star(\beta) W(\beta)V(\alpha)}
= \frac{p(\beta, \alpha)}{p(\alpha, \beta)}.
\end{equation}
This parameter is equal  to the ratio of the ordered joint probabilities of the same outcome,
but in the different order, namely,  ``$b$ then $a$'' or ``$a$ then $b$''. 
Then \cite{LOUB}
\begin{equation}
\label{EQR12}
 \lambda_\beta= \frac{1}{2} \Big[ \sqrt{\frac{p(\alpha_1, \beta)}{p(\alpha_2, \beta)}} (\gamma_{\alpha_1 \beta} -1) 
 + \sqrt{\frac{p(\alpha_2, \beta)}{p(\alpha_1, \beta)}} (\gamma_{\alpha_2 \beta} -1)
 \Big].
 \end{equation}
In principle, this coefficient can be larger than one \cite{LOUB}. Hence, it cannot be represented as 
\begin{equation}
\label{EQR12a}
\lambda_\beta= \cos \theta_\beta
\end{equation}
 for some angle (``phase'') $\theta_\beta,$ cf. (\ref{AS}).
However, if POVMs $M^a$ and $M^b$ are, in fact, spectral decompositions of Hermitian operators, 
then the coefficients of interference are always less than one, i.e., one can find phases 
$\theta_\beta.$ 

The transition from the classical FTP to the FTP with the interference term (\ref{EQR9}) can be considered
as an extension of the parameter space. Besides the probabilities for the results $a=\alpha$ and $b=\beta$  and 
the conditional probabilities $p(\beta \vert \alpha),$ new parameters $\lambda_\beta,$ the coefficients of interference
between observables $a$ and $b$ (for the state $\rho),$ are invented in consideration. 

We present an interpretation of the structure of the coefficients of interference. We start with terms $\gamma_{\alpha \beta}.$
They express {\it noncommutativity of measurements} or nonclassicality of the joint probability distribution. 
For classical probability all coefficients 
$\gamma_{\alpha \beta} =1$ (and hence all interference coefficients $\lambda_{\beta}=0).$ 

We also introduce the coefficients 
\begin{equation}
\label{EQR12bb}
\mu_\beta = \frac{p(\alpha_1, \beta)}{p(\alpha_2, \beta)}.
 \end{equation}
 They express the relative magnitude of probabilistic influences of the results $a=\alpha_1$  and $a=\alpha_2,$ respectively, on the result 
 $b=\beta.$
 
 Thus the coefficients of interference are composed of the coefficients of noncommutativity which are wighted with the 
 the relative magnitudes of influences:
\begin{equation}
\label{EQR12bb1}
\lambda_\beta= \frac{1}{2} [ (\gamma_{\alpha_1 \beta} -1) \sqrt{\mu_\beta} +   (\gamma_{\alpha_2 \beta} -1)/ \sqrt{\mu_\beta}].
\end{equation}

\section{The scheme of decision  making based on classical and quantum-like FTP}
\label{DM}
\subsection{Classical scheme}

There are two parties, Alice and Bob. They operate under some complex of conditions
(physical, social, financial) $C,$ context. Each can act only in two ways: $a=\alpha_1, \alpha_2$ and 
$b=\beta_1, \beta_2.$  Decisions of Bob depend crucially on  actions of Alice. However, in general 
Bob does not know precisely which action $a=\alpha_1$ or $a= \alpha_2$ will be chosen by Alice, Therefore 
Bob can only guess. He estimates subjectively probabilities $p^a(\alpha_1)$ and  $p^a(\alpha_2)$
of possible actions of Alice. He can also estimate probabilities of his own actions conditioned 
on actions of Alice, $p(\beta\vert \alpha).$ Finally, Bob estimates the probabilities 
$p^b(\beta_1)$ and $p^b(\beta_2)$ by using FTP (\ref{EQR8}). If, e.g.,  $p^b(\beta_1) >p^b(\beta_2),$
then Bob takes the decision $b=\beta_1.$

This scheme is realized in the brain on the unconscious level. Subjective estimates of probabilities
for actions of Alice and conditional probabilities for his own actions conditioned by the actions of Alice
typically not present on the conscious level. 
 
\subsection{Quantum-like scheme}

This is a generalization of classical scheme. Besides the aforementioned probabilities, Bob estimates 
the measure of incompatibility of his possible actions and possible actions of Alice, the coefficients 
of interference. The later estimate is combined  of the estimate of the effect of noncommutativity of Bob's and Alice's 
actions and relative magnitudes of probabilistic influences of Alice's actions to Bob's actions. 
To estimate $\gamma_{\alpha \beta},$ Bob has to estimate not only the ordered probability $p(\alpha, \beta)$
of his own action $b=\beta$ under the condition that Alice would act as $a=\alpha,$ but  also 
the ordered probability $p(\beta, \alpha)$ of possible Alice's action $a=\alpha$ under the condition 
that she assumes Bob's action $b=\beta.$ However, the absolute magnitudes 
of these probabilities are not important, Bob is interested only in their ratio.

Finally, Bob estimates probabilities of his actions by using the FTP with the interference term (\ref{EQR9}).

This scheme is also realized on the unconscious level. In particular, all aforementioned measures of quantumness
are estimated subjectively (on the basis of collected experience).

We can say that in the quantum-like scheme the counterfactual arguments play a crucial role. To estimate
the coefficients of noncommutativity $\gamma_{\alpha \beta},$ Bob has to imagine how Alice woudl act 
if he acts as $b=\beta.$ These probabilities are compared with probabilities of Bob's own actions 
condtioned by possible actions of Alice. If Bob guess that Alice would make the same estimation of the 
probabilities of his actions  conditioned by her actions as for her own actions conditioned 
the corresponding actions of Bob, then $p(\alpha, \beta) = p(\beta, \alpha)$ and $\gamma_{\alpha \beta}=1.$ 
In this case the quantum-like and classical FTP and, hence, decision making schemes coincide.
Thus {\it in this decision making scheme quantumness is related assymetry in the estimation of the action-reaction 
probabilities.} Bob can imagine that Alice is frendlier than him or that Alice is more defect inclined
(we consider the Prisoner's Dilemma game as the basic example of the situation under consideration). 

In the operational quantum-like formalism all probabilities and components of the coefficients of interference
are encoded in the Bob's mental state, the density operator $\rho,$ his decision making observable, POVM $M^b,$ and 
his observable for intentions of Alice, POVM $M^�.$ (We remark these are self-observables.) 
    
\section*{Acknowledgments} 

Research of one of the authors (Irina Basieva) was based on the post-doc fellowship at Linnaeus University (the grant ``Mathematical 
Modeling of Complex Hierarchic Systems'').

\end{document}